\documentclass[seceq]{ptptex}

\usepackage{amsmath}
\usepackage{amssymb}
\usepackage{bm}
\usepackage[dvips]{graphicx}

\newcommand{\N}{{\mathcal N}}

\newcommand{\V}{{\mathcal V}}

\newcommand{\Tr}{\text{Tr}}
\newcommand{\tr}{\text{tr}}
\newcommand{\Str}{\text{Str}}

\preprintnumber[3cm]{her-th/0512103\\  KUNS-1999\\ Dec, 2005}

\title{
Super Yang-Mills Theory from a Supermatrix Model %
}

\author{
Takeshi  \textsc{Morita}\footnote{
E-mail: tmorita@post.kek.jp} %
}

\inst{
Theory Group, KEK, Tsukuba 305-0801, Japan
}

\abst{

It is known that Yang-Mills theories on non-commutative space can be derived from large-$N$ reduced models.
Gauge fields in non-commutative Yang-Mills theories can be described as fluctuations of matrices expanded about an appropriate classical solution of the reduced models.
We investigate a generalization of this procedure in superfield formalism.
We show that we can construct a supermatrix model such that $D=4$ $\N=1$ super Yang-Mills theory can be derived from it.
In addition, we can couple matter supermatrices to this supermatrix model and also construct models corresponding to $\N=2$ and $\N=4$ super Yang-Mills theories.
In these investigations, we need to introduce a new non-anti-commutative superspace, and we investigate the definition of field theories on this space.

}

\begin{document}

\maketitle

\section{Introduction}
\setcounter{equation}{0}
\setcounter{footnote}{0}

Matrix models\cite{IKKT,BFSS,Smolin} represent important proposals for a non-perturbative definition of string theory.
Through various studies, it has been revealed that the matrices employed in these models can describe gauges and fields \cite{EK}, space-time\cite{NCYM, HKK}, D-branes\cite{IKKT} and strings\cite{DVV, Wilson}.

Here we explain the derivation of Yang-Mills theory on non-commutative space\cite{SW, Wess} from matrix models.\cite{NCYM} \
(Further explanation and definitions of our notation are in Appendix \ref{a1}.)
We consider the large-$N$ reduced model of Yang-Mills theory, whose action is given by
\begin{align}
S=\frac{1}{g^2} \Tr_{U(\hat{N})} \left(\frac{1}4 [\hat{A}_\mu, \hat{A}_\nu]^2 \right).
\label{IKKT}
\end{align}
This action has a classical solution $\hat{A}_\mu=\hat{p}_\mu$, corresponding to a derivative operator on a non-commutative space.
Considering fluctuations of $\hat{A}_\mu$ expanded about this solution, we can obtain the action of Yang-Mills theory on the non-commutative space,
\begin{align}
S=  \int d^{d} x \frac{1}{{\tilde{g}}^2}  \tr \left(-\frac{1}4 F_{\mu\nu}^2  \right)_*,\label{NCYM}
\end{align}
where $*$ indicates that the product of this action is the $*$-product, defined in (\ref{star*}). 

This derivation is quite interesting.
The original reduced model, (\ref{IKKT}), is a 0-dimensional theory.
However, the derived non-commutative Yang-Mills theory, (\ref{NCYM}), has coordinate dependence.
This suggests that non-commutative space can be described as a configuration of the matrices.
In fact, as we show in Appendix \ref{a1}, translation of non-commutative coordinates, $x^\mu \rightarrow x^\mu + \epsilon^\mu$, can be regarded as a gauge transformation in the reduced model.
Generally, such a procedure can be regarded as the generation of non-commutative space from the matrix models.\\

In this paper, we apply this study to a derivation of super Yang-Mills on non-commutative space from a supermatrix model.
We construct a new $GL(2|2,R)\times U(\hat{N})$ supermatrix model.
The action of this supermatrix model is given by
\begin{align}
S=\frac{\hat{N}}{g_m}2\pi i \tau \Str \left(
 \{ \hat{A}_\alpha, \bar{\hat{A}}_{\dot{\alpha}}  \}^2  \right).
\label{SMM1}
\end{align}
We apply the above described procedure to derive super Yang-Mills from this model.
This model has a classical solution $\hat{A}_\alpha = \hat{\pi}_\alpha$ corresponding to a derivative operator on a non-commutative superspace.\footnote{
In precise terms, fermionic coordinates $\theta^\alpha$ and $\bar{\theta}^{\dot{\alpha}}$  become non-anti-commutative. 
However, we call them `non-commutative' fermionic coordinates for simplicity.
In this paper, we consider three types of non-commutative coordinates.
These constitute the `non-commutative superspace' $(\theta, \bar{\theta})$, the `non-commutative space' $(x^\mu)$ and the `4-dimensional non-commutative superspace' $(x^\mu, \theta, \bar{\theta} )$.}\cite{S, OV1, OV2, KKM, NCSS, ST} \
Considering fluctuations of $\hat{A}_\alpha$ expanded about this solution, we can obtain the action of a reduced model on the non-commutative superspace,
\begin{align}
S= \frac{\hat{N}}{\tilde{g}_m} \int d^2\theta ~2\pi i \tau\Tr_{U(\hat{N})}
\left(-\frac{1}4   \frac{\partial}{\partial \bar{\theta}^{\dot{\alpha}}} \frac{\partial}{\partial \bar{\theta}_{\dot{\alpha}}} e^{-\hat{\V}} \frac{\partial}{\partial \theta^\alpha} e^{\hat{\V}} \right)^2_\star + \text{c.c.},
\label{SYM5}
\end{align}
where $\star$ indicates that the product of this action is the $\star$-product, which we define in subsequent section.
Here $\hat{\V}$ is an $\hat{N} \times \hat{N}$ matrix whose matrix elements are function of the non-commutative coordinates $\theta$ and $\bar{\theta}$.
Thus, this procedure can be regarded as a generation of the non-commutative superspace from the supermatrix model. 
This model is a large-$N$ reduced model of $D=4$, $\N=1$ super Yang-Mills theory on the non-commutative superspace.
In fact, if we ignore the $x^\mu$ dependence in $D=4$, $\N=1$ $U(\hat{N})$ super Yang-Mills theory in the superfield formalism, we can obtain (\ref{SYM5}) in the usual anti-commutative $\theta$, $\bar{\theta}$ space.

From (\ref{SYM5}), if we take the commutative limit of the non-commutative fermionic coordinates, we can derive the action of super Yang-Mills theory\footnote{We use the notation of the textbook by J. Wess and J. Bagger\cite{BW}.} on non-commutative space
\begin{align}
S=\int d^4x d^2 \theta 2\pi i \tau \Tr_{U(n)}  \left( -\frac{1}{4}\bar{D}_{\dot{\alpha}} \bar{D}^{\dot{\alpha}} e^{-V} D_\alpha e^V \right)_*^2+\text{c.c.}.
\label{SYM}
\end{align}
Consequently, we can obtain the super Yang-Mills theory from the supermatrix model.

This result is also interesting.
Though the original supermatrix model (\ref{SMM1}) does not have superspace $\theta$, $\bar{\theta}$ dependence, the reduced model (\ref{SYM5}) does.
This implies that superspace can also be described as a configuration of the supermatrices.
\\

This paper is organized as follows.
In \S \ref{2.0}, we study the derivation of super Yang-Mills theory.
In \S \ref{2.3}, we show that we can couple matter supermatrices to the supermatrix model.
Therefore, we can study the relation to the Dijkgraaf-Vafa theory.\cite{DV} \ 
In addition, we can construct supermatrix models corresponding to $\N=2$ and $\N=4$ super Yang-Mills theories.
Section \ref{Discussions} is devoted to discussion and the proposal of possible applications.
In our derivation of super Yang-Mills theory, we need to introduce a new non-commutative superspace which is different from that used in most studies.\cite{S} \
We also investigate the construction of field theories on this superspace.

Appendix \ref{a1} contains a review of the relation between the non-commutative Yang-Mills theory and the large-$N$ reduced model.
In Appendix \ref{a2}, we review the construction of $\N=1$ super Yang-Mills theory through the covariant approach.\cite{superspace} \ 
This approach is important in the study of our supermatrix model.
In Appendix \ref{a3}, we present an explicit expression of our non-commutative coordinates in terms of $GL(2|2,R)$ matrices.

\section{Super Yang-Mills theory from a supermatrix model}
\setcounter{equation}{0}
\label{2.0}
The main purpose of this paper is to derive the reduced model (\ref{SYM5}) and the super Yang-Mills theory (\ref{SYM}) from the supermatrix model (\ref{SMM1}).
We explain the details of this supermatrix model in \S \ref{2.1}.
In \S \ref{3.1}, we introduce our new non-commutative superspace and discuss its relation to the supermatrix model.
Using this relation, we study the derivation of the super Yang-Mills in \S \ref{2.2}.

\subsection{Supermatrix model}
\label{2.1}

We investigate a $GL(2|2,R)\times U(\hat{N})$ supermatrix model.
Here we take $\hat{N}$ to be infinite.
The action of this model is given by
\begin{align}
S=\frac{\hat{N}}{g_m}2\pi i \tau \Str \left(
 \epsilon^{\alpha \beta} \epsilon_{\dot{\alpha} \dot{\beta}} 
 \{ \hat{A}_\alpha, \bar{\hat{A}}^{\dot{\alpha}} \}
 \{ \hat{A}_\beta, \bar{\hat{A}}^{\dot{\beta}} \}  \right),
\label{SMM}
\end{align}
where $g_m$ is a constant, its mass dimension is 2, and $\tau= 4\pi i/g^2 $ is the gauge coupling constant.\footnote{ More generally, $\tau$ can be taken as $\tau= \theta/2 \pi+ 4\pi i/g^2 $.
However, it is difficult to treat the $\theta$ term in reduced models, because $ \Tr F_{\mu\nu} \tilde{F}^{\mu\nu} \rightarrow \Tr \left(\epsilon^{\mu\nu\rho\sigma}  [\hat{A}_\mu,\hat{A}_\nu] [\hat{A}_\rho, \hat{A}_\sigma] \right)=0  $. Therefore we do not consider this term in this paper.}
The spinor indices ($\alpha, \dot{\alpha}=1,2$) are contracted by the anti-symmetric tensor $\epsilon_{\alpha \beta}$ and $\epsilon_{\dot{\alpha}\dot{\beta}}$, and this model has $SL(2,C)$ symmetry.
The quantities $\hat{A}_\alpha$ are ferimionic $GL(2|2,R)\times U(\hat{N})$ matrices and are defined by $\hat{A}_\alpha=A^a_{i\alpha} T^a \otimes t^i $ , where $T^a$ $(a=1\ldots \hat{N}^2)$ are the generators of $U(\hat{N})$ and $t^i$ $(i=1\ldots 16)$ are those of $GL(2|2,R)$.
The matrices $\bar{\hat{A}}_{\dot{\alpha}}$ are the hermite conjugates of $\hat{A}_\alpha$.
Both matrices have mass dimension $1/2$, corresponding to that of $\partial/\partial \theta$ and  $\partial/\partial \bar{\theta}$.

The definition of the supertrace is 
\begin{align}
\Str (\hat{O}) \equiv \Tr \left( 
\begin{pmatrix}
 1_{2\hat{N}}&\\
 &-1_{2\hat{N}}
 \end{pmatrix}  
 \hat{O}
    \right),
\label{supertrace}
\end{align}
where $1_{2\hat{N}}$ is a $2\hat{N} \times 2\hat{N}$ identity matrix.

The model is invariant under  $GL(2|2,R)\times U(\hat{N})$ gauge transformations, and hence we have
\begin{align}
\hat{A}'_\alpha= e^{-i\hat{K}} \hat{A}_\alpha  e^{i\hat{K}},~
\bar{\hat{A}}'_{\dot{\alpha}}= e^{-i\hat{K}} \bar{\hat{A}}_{\dot{\alpha}}  e^{i\hat{K}},
\label{symmetry}
\end{align}
where $\hat{K}=K^a_{i} T^a \otimes t^i $, which satisfies $\hat{K}=\bar{\hat{K}}$.

Furthermore, we impose constraints
\begin{align}
\{ \hat{A}_\alpha, \hat{A}_\beta \} = \{ \bar{\hat{A}}_{\dot{\alpha}} , \bar{\hat{A}}_{\dot{\beta}} \}=0
\label{constraint2}
\end{align}
on this model.
The meaning of these constraints will be explained in \S \ref{2.2}.

\subsection{Non-commutative superspace}
\label{3.1}

We now introduce a non-commutative superspace and discuss a relation to the $GL(2|2,R)$ matrix in this subsection.
This relation is similar to that between the non-commutative space and the $U(\hat{N})$ matrix.

We consider fermionic non-commutative coordinates, which satisfy the (anti-)commutator relations
\begin{align}
\{ \hat\theta^{\alpha} , \bar{\hat{\theta}}^{\dot{\alpha}} \} = \gamma^{\alpha \dot{\alpha}} , \nonumber \\
\{ \hat\theta^\alpha , \hat\theta^\beta \}= \{ \bar{\hat{\theta}}^{\dot{\alpha}} , \bar{\hat{\theta}}^{\dot{\beta}} \} =0,
\label{NCSS2}
\end{align}
where $\gamma^{\alpha \dot{\alpha}}$ are c-numbers.
This non-commutativity is different from those studied in Ref. \citen{S} and Ref. \citen{KKM}. 
In \S \ref{2.4}, we explain why we choose it and discuss the construction of 4-dimensional field theories on such a space.

This non-commutative superspace can be represented by $ GL(2|2,R) $ matrices.
It is convenient to take $\gamma^{\alpha \dot{\alpha}}=  \gamma \delta^{\alpha \dot{\alpha}}$ by applying a $SL(2,C)$ transformation.
Then, the following matrices satisfy the (anti-)commutator relations (\ref{NCSS2}):
\begin{align}
\hat\theta^1 = \sqrt\gamma 
\begin{pmatrix}
0&0&0&1\\
0&0&0&0\\
0&-1&0&0\\
0&0&0&0
\end{pmatrix}
 ,~\hat\theta^2 =  \sqrt{\gamma}
\begin{pmatrix}
0&0&1&0\\
0&0&0&0\\
0&0&0&0\\
0&1&0&0
\end{pmatrix}
\nonumber,  \\
\bar{\hat{\theta}}^{\dot 1} = \sqrt{\gamma}
\begin{pmatrix}
0&0&0&0\\
0&0&-1&0\\
0&0&0&0\\
1&0&0&0
\end{pmatrix}
,~\bar{\hat{\theta}}^{\dot 2} = \sqrt{\gamma}
\begin{pmatrix}
0&0&0&0\\
0&0&0&1\\
1&0&0&0\\
0&0&0&0
\end{pmatrix}.
\end{align}
Note that these matrices, their products $\hat{\theta}^2$, $\hat{\theta}^\alpha \bar{\hat{\theta}}^{\dot{\alpha}}$, $\dots$, $\hat{\theta}^2 \bar{\hat{\theta}}^2$ and the identify matrix $1_4$ are all linearly independent.
In Appendix \ref{a3}, this is confirmed by studying their explicit expressions.
Therefore, $1_4$, $\hat{\theta}$, $\cdots$, $\hat{\theta}^2 \bar{\hat{\theta}}^2$ can be regarded as the bases of $GL(2|2,R)$.

We can regard these matrices as fermionic non-commutative coordinates, and consider a field theory on them corresponding to the matrix model.\cite{KKM, NCSS, ST, JP} \
First, a matrix $\hat{O}$ can be mapped to a corresponding field $O(\theta, \bar{\theta})$ on the non-commutative superspace via the Weyl ordering:
\begin{align}
\hat{O}= \int 2^4 d^2 \kappa d^2\bar{\kappa} ~ \tilde{O}(\kappa, \bar{\kappa}) 
e^{\kappa^\alpha \hat{\theta}_\alpha + \bar{\kappa}_{\dot{\alpha}} \bar{ \hat{\theta}}^{\dot{\alpha}} } 
 \longleftrightarrow
O(\theta, \bar{\theta})= \int 2^4 d^2 \kappa d^2\bar{\kappa} ~ \tilde{O}(\kappa, \bar{\kappa}) 
e^{\kappa^\alpha \theta_\alpha + \bar{\kappa}_{\dot{\alpha}} \bar{\theta}^{\dot{\alpha}} }
 \label{Weyl}
\end{align}
Then, the product of two matrices, $\hat{O}_1 \hat{O}_2$, corresponds to $O_1\star O_2(\theta, \bar{\theta})$, where this $\star$-product is defined by
\begin{align}
O_1\star O_2(\theta , \bar{\theta} )= \left. \exp \left( -\frac{1}{2} \gamma^{\alpha \dot{\alpha}} \left( \frac{\partial}{\partial \theta_1^\alpha}
  \frac{\partial}{\partial \bar{\theta}_2^{\dot{\alpha}}  }
  + \frac{\partial}{\partial \bar{\theta}_1^{ \dot{\alpha}}}
  \frac{\partial}{\partial \theta_2^{\alpha}  } \right)  \right)
   O_1(\theta_1 , \bar{\theta}_1 )O_2(\theta_2 , \bar{\theta}_2 ) \right|_{\theta=\theta_1=\theta_2 }.\label{star}
\end{align}

Next, we introduce derivative operators on this space as
\begin{align}
\hat{\pi}_{\alpha}\equiv \beta_{\alpha \dot{\alpha}} \bar{\hat{\theta}}^{\dot{\alpha}},~\bar{\hat{\pi}}_{\dot{\alpha}} \equiv \beta_{\alpha \dot{\alpha}}
\hat{\theta}^\alpha,~~\hat{\pi}_{\alpha}\equiv \beta_{\alpha \dot{\alpha}} \bar{\hat{\theta}}^{\dot{\alpha}},~\bar{\hat{\pi}}_{\dot{\alpha}} \equiv \beta_{\alpha \dot{\alpha}}
\hat{\theta}^\alpha,
\label{derivative2}
\end{align}
where $\beta_{\alpha \dot\beta}$ is the inverse of $\gamma^{\alpha \dot\beta}$:
\begin{align}
\gamma^{\alpha\dot{\alpha}}\beta_{\dot{\alpha} \beta}= \delta^{\alpha}_{\beta},~~
\beta_{\dot{\alpha} \alpha} \gamma^{\alpha \dot{\beta}}= \delta_{\dot{\alpha}}^{\dot{\beta}}.
\end{align}
Then, these matrices satisfy the following anti-commutation relations
\begin{align}
&\{ \hat{\pi}_{\alpha} , \hat{\theta}^\beta \} = \delta_\alpha^\beta,~
\{ \bar{ \hat{\pi}}_{\dot{\alpha}}, \bar{\hat{\theta}}^{\dot{\beta}} \} = \delta_{\dot{\alpha}}^{\dot{\beta}},~ 
\{ \hat{\pi}_{\alpha}, \bar{ \hat{\pi}}_{\dot{\alpha}} \}= \beta_{\alpha \dot{\alpha}}, \nonumber \\
&\{ \hat{\pi}_{\alpha}, \bar{\hat{\theta}}^{\dot{\alpha}} \} =
\{  \bar{ \hat{\pi}}_{\dot{\alpha}}, \hat{\theta}^\alpha \} =
\{ \hat{\pi}_{\alpha}, \hat{\pi}_{\beta} \} =
\{ \bar{ \hat{\pi}}_{\dot{\alpha}}, \bar{ \hat{\pi}}_{\dot{\beta}} \}=0.
\end{align}
Hence $\{ \hat{\pi}_\alpha ,\hat{O} ]$ corresponds to $\partial /\partial \theta^\alpha O $, 
where we take a commutator or an anti-commutator according to the statistics of $\hat{O}$. 

Finally, an integral on this space is equivalent to a supertrace:
\begin{align}
\int d^2 \theta d^2 \bar{\theta}~ O =
 \frac{1}{4 \gamma^2} \Tr_{GL(4,R)} \left( 
 \begin{pmatrix}
 1_{2}&\\
 &-1_{2}
 \end{pmatrix}  
     \hat{O} \right)
\equiv  \frac{1}{4\det \gamma} \Str \hat{O}.
\label{supertrace5}
\end{align}
Here, the mass dimension of $\det \gamma$ is $-2$, which is consistent with the dimension of the left-hand side.

Through these relations, we can map $GL(2|2,R)$ matrix models to field theories on the non-commutative superspace.

\subsection{Super Yang-Mills theory from the supermatrix model}
\label{2.2}
In the first half of this subsection, we derive the reduced model (\ref{SYM5}) on the non-commutative superspace from the supermatrix model (\ref{SMM}).
In the second half, we consider the commutative limit of this superspace and show that $D=4$ $\N=1$ super Yang-Mills theory on non-commutative {\it space} (\ref{SYM}) can be derived from the reduced model (\ref{SYM5}).\cite{KKM} \\

We start by analyzing the supermatrix model (\ref{SMM}).
Because we impose the constraints (\ref{constraint2}) on the supermatrix model, the action becomes
\begin{align}
S=\frac{\hat{N}}{g_m}2\pi i \tau \Str \left(
 \epsilon^{\alpha \beta} \epsilon_{\dot{\alpha} \dot{\beta}} 
 \{ \hat{A}_\alpha, \bar{\hat{A}}^{\dot{\alpha}} \}
 \{ \hat{A}_\beta, \bar{\hat{A}}^{\dot{\beta}} \} 
 + \hat{\Lambda}^{\alpha\beta} \{ \hat{A}_\alpha, \hat{A}_\beta \}
 + \bar{\hat{\Lambda}}_{\dot{\alpha} \dot{\beta}} \{  \bar{\hat{A}}^{\dot{\alpha}},
 \bar{\hat{A}}^{\dot{\beta}}  \}  \right),
\end{align}
where $\hat{\Lambda}_{\alpha\beta}$ and $\bar{\hat{\Lambda}}_{\dot{\alpha} \dot{\beta}} $ are Lagrange multipliers.
Then, we can derive the equations of motion,
\begin{align}
[ \bar{\hat{A}}_{\dot{\alpha}} , \{ \bar{\hat{A}}^{\dot{\alpha}} , \hat{A}^\alpha \}]
+[\hat{\Lambda}^{\alpha\beta},\hat{A}_\beta  ]&=0, \nonumber \\
[ \hat{A}^\alpha , \{ \hat{A}_\alpha , \bar{\hat{A}}_{\dot{\alpha}} \}] +
[ \bar{\hat{\Lambda}}_{\dot{\alpha}\dot{\beta} },\bar{\hat{A}}^{\dot{\beta}} ]&=0,\nonumber \\
\{ \hat{A}_\alpha, \hat{A}_\beta \} = \{ \bar{\hat{A}}_{\dot{\alpha}} , \bar{\hat{A}}_{\dot{\beta}} \}&=0,
\label{eom}
\end{align}
and we can find the classical solution 
\begin{align}
\hat{A}_\alpha = \hat{\pi}_\alpha \otimes 1_{\hat{N}},~
\bar{\hat{A}}_{\dot{\alpha}} = \bar{\hat{\pi}}_{\dot{\alpha}} \otimes 1_{\hat{N}},~
\hat{\Lambda}_{\alpha\beta}=\bar{\hat{\Lambda}}_{\dot{\alpha}\dot{\beta}}=0,
\label{solution5}
\end{align}
where $\hat{\pi}_\alpha$ is defined in (\ref{derivative2}).

We consider fluctuations of $\hat{A}_\alpha$ expanded about this solution.
The constraints (\ref{constraint2}) require these fluctuations to be
\begin{align}
\hat{A}_\alpha=e^{-\hat{\omega}} \hat{\pi}_\alpha e^{\hat{\omega}},~
\bar{\hat{A}}_{\dot{\alpha}}=e^{\bar{\hat{\omega}}} \bar{\hat{\pi}}_{\dot{\alpha}} e^{-\bar{\hat{\omega}}},
\label{solution}
\end{align}
where $\hat{\omega}$ is an arbitrary complex matrix.
Here, the expression $\hat{A}_\alpha=e^{-\hat{\omega}} \hat{\pi}_\alpha e^{\hat{\omega}}$ is reminiscent of the expression of the covariant derivative $\nabla_{\alpha}=e^{-\Omega} D_\alpha e^{\Omega}$ in the covariant approach (\ref{covariant}).
In the case of the derivation of non-commutative Yang-Mills theory, it is known that the matrices $\hat{A}_{\mu}$ correspond to the covariant derivative $D_\mu=\partial_\mu + i a_\mu$, where $a_\mu$ are the gauge fields.
In our case, the matrices $\hat{A}_\alpha$ correspond to $\nabla_\alpha$.\footnote{
Precisely, the matrices $\hat{A}_\alpha$ correspond to $ e^{-\hat{\Omega}} \partial/\partial \theta^\alpha e^{\hat{\Omega}}$, which are covariant derivatives in the reduced model (\ref{SYM6}). 
These $ e^{-\hat{\Omega}} \partial/\partial \theta^\alpha e^{\hat{\Omega}}$ correspond to $\nabla_\alpha$ when we take the commutative limit.}
Accordingly, the constraints (\ref{constraint2}) correspond to the representation preserving constraints (\ref{constraint1}).

Then, the action expanded about the solution becomes
\begin{align}
S=\frac{\hat{N}}{g_m}2\pi i \tau \Str \left(  \{ e^{-\hat{\omega}} \hat{\pi}_\alpha e^{\hat{\omega}}, e^{\bar{\hat{\omega}}} \bar{\hat{\pi}}_{\dot{\alpha}} e^{-\bar{\hat{\omega}}} \}^2
  \right).
\end{align}
Through the relations studied in \S \ref{3.1}, we can map this action to
\begin{align}
S=\frac{\hat{N}}{g_m} 4\det{\gamma} \int d^2\theta d^2 \bar{\theta} ~2\pi i \tau 
\Tr_{U(\hat{N})} \left(  \left( \left\{ e^{-\hat{\Omega}} \frac{\partial}{\partial \theta^\alpha} e^{\hat{\Omega}}, e^{\bar{\hat{\Omega}}}  \frac{\partial}{\partial \bar{\theta}^{\dot{\alpha}}} e^{-\bar{\hat{\Omega}}} \right\} + \beta_{\alpha \dot{\alpha}}       \right)^2 \right)_\star,
\label{SYM6}
\end{align}
where the trace is taken over the $U(\hat{N})$ group.
The matrices $\hat{\omega}$ has been mapped to $\hat{\Omega}(\theta, \bar{\theta})$ by the Weyl ordering (\ref{Weyl}), and this $\hat{\Omega}$ is an $\hat{N}\times \hat{N}$ matrix whose elements are functions of the non-commutative coordinates $\theta$ and $\bar{\theta}$.

Here we emphasize that the original supermatrix model (\ref{SMM}) does not have $\hat{\theta}$ dependence.
By considering this model expanded about the classical solution (\ref{solution5}), $\hat{\theta}$ is introduced.
Consequently, we can map the supermatrix model to the reduced model (\ref{SYM6}).
This is the generation of the non-commutative superspace from the supermatirx model.

Let us now derive the action (\ref{SYM5}). 
First, we expand the term in the parentheses of (\ref{SYM6}).
This gives
\begin{align*}
S=&\frac{\hat{N}}{g_m} 4\det{\gamma} \int d^2\theta d^2 \bar{\theta} ~2\pi i \tau 
\Tr \left(   \left\{ e^{-\hat{\Omega}} \frac{\partial}{\partial \theta^\alpha} e^{\hat{\Omega}}, e^{\bar{\hat{\Omega}}}  \frac{\partial}{\partial \bar{\theta}^{\dot{\alpha}}} e^{-\bar{\hat{\Omega}}} \right\}^2 \right.  \\
&\left. + 2\beta^{\alpha \dot{\alpha}} \left\{ e^{-\hat{\Omega}} \frac{\partial}{\partial \theta^\alpha} e^{\hat{\Omega}}, e^{\bar{\hat{\Omega}}}  \frac{\partial}{\partial \bar{\theta}^{\dot{\alpha}}} e^{-\bar{\hat{\Omega}}} \right\}   +  (\beta^{\alpha \dot{\alpha}} )^2  \right) _\star.
\end{align*}
Here we can ignore the second and the third terms, since the second is proportional to the trace of the anti-commutator and the third is a constant factor. 
Then, the action can be calculated as
\begin{align*}
S=&\frac{\hat{N}}{g_m} 4 \det{\gamma} \int d^2\theta d^2 \bar{\theta} ~2\pi i \tau 
\Tr   \left( \{ e^{-\hat{\Omega}} \frac{\partial}{\partial \theta^\alpha} e^{\hat{\Omega}}, e^{\bar{\hat{\Omega}}}  \frac{\partial}{\partial \bar{\theta}^{\dot{\alpha}}} e^{-\bar{\hat{\Omega}}} \}^2     \right)_\star \nonumber \\
=&2\frac{\hat{N}}{g_m} \det{\gamma} \int d^2\theta d^2 \bar{\theta} ~2\pi i \tau\Tr   \left(e^{-\bar{\hat{\Omega}}} \{ e^{-\hat{\Omega}} \frac{\partial}{\partial \theta^\alpha} e^{\hat{\Omega}}, e^{\bar{\hat{\Omega}}}
  \frac{\partial}{\partial \bar{\theta}^{\dot{\alpha}}} e^{-\bar{\hat{\Omega}}} \}   e^{\bar{\hat{\Omega}}}  \right)^2_\star \nonumber \\
&+2\frac{\hat{N}}{g_m} \det{\gamma} \int d^2\theta d^2 \bar{\theta} ~2\pi i \tau\Tr   \left( e^{\hat{\Omega}} \{ e^{-\hat{\Omega}} \frac{\partial}{\partial \theta^\alpha} e^{\hat{\Omega}}, e^{\bar{\hat{\Omega}}}  \frac{\partial}{\partial \bar{\theta}^{\dot{\alpha}}} e^{-\bar{\hat{\Omega}}} \}    e^{-\hat{\Omega}} \right)^2_\star.
\end{align*}
In the last equation, every $\hat{\Omega}$ and $\bar{\hat{\Omega}}$ appear in the combination $ e^{\hat{\Omega}} e^{\bar{\hat{\Omega}}} $ or its inverse, and thereby it is convenient to define
\begin{align}
e^{\hat{\V}} \equiv e^{\hat{\Omega}} e^{\bar{\hat{\Omega}}}.
\end{align}
From this definition, it is clear that $\hat{\V}$ satisfies $\bar{\hat{\V}} = \hat{\V}$, and thus $\hat{\V}$ corresponds to a vector superfield.
By using this, the action becomes
\begin{align}
S=& 2\frac{\hat{N}}{g_m} \det{\gamma} \int d^2\theta d^2 \bar{\theta} ~2\pi i \tau\Tr   \left(
 \frac{\partial}{\partial \bar{\theta}^{\dot{\alpha}}} e^{-\hat{\V}} \frac{\partial}{\partial \theta^\alpha} e^{\hat{\V}} \right)^2_\star \nonumber \\
&+2\frac{\hat{N}}{g_m} \det{\gamma} \int d^2\theta d^2 \bar{\theta} ~2\pi i \tau\Tr  
 \left(  \frac{\partial}{\partial \theta^\alpha} e^{\hat{\V}}  \frac{\partial}{\partial \bar{\theta}^{\dot{\alpha}}} e^{-\hat{\V}}  \right)^2_\star .
\label{action3} 
\end{align}
As is well known, the integral of the fermionic coordinates can be replaced by a derivative as
\begin{align}
\int d^2 \bar{\theta} = -\frac{1}{4} \epsilon^{\dot{\alpha} \dot{\beta}}  \frac{\partial}{\partial \bar{\theta}^{\dot{\alpha}}} \frac{\partial}{\partial \bar{\theta}^{\dot{\beta}}}.
\label{integral}
\end{align}
Then, we obtain the action
\begin{align}
S= & 8\frac{\hat{N}}{g_m} \det{\gamma} \int d^2\theta ~2\pi i \tau\Tr
\left(-\frac{1}4   \frac{\partial}{\partial \bar{\theta}^{\dot{\alpha}}} \frac{\partial}{\partial \bar{\theta}_{\dot{\alpha}}} e^{-\hat{\V}} \frac{\partial}{\partial \theta^\alpha} e^{\hat{\V}} \right)^2_\star \nonumber \\
&+8\frac{\hat{N}}{g_m} \det{\gamma} \int  d^2 \bar{\theta} ~2\pi i \tau\Tr 
\left(-\frac{1}4  
  \frac{\partial}{\partial \theta^\alpha}\frac{\partial}{\partial \theta_\alpha}
  e^{\hat{\V}}  \frac{\partial}{\partial \bar{\theta}^{\dot{\alpha}}} e^{-\hat{\V}}  \right)^2_\star.
\label{action4}
\end{align}
This action represents the large-$N$ reduced model of $D=4$, $\N=1$ super Yang-Mills theory on the non-commutative superspace.
Actually, when we take the commutative limit $\gamma^{\alpha\dot{\alpha}} \rightarrow 0$, we can obtain this model by simply ignoring the $x^\mu$ dependence in the $U(\hat{N})$ super Yang-Mills theory in the superfield formalism.
\\

In the latter half of this subsection, we take the commutative limit $\gamma^{\alpha\dot{\alpha}} \rightarrow 0$ and derive the super Yang-Mills theory from the reduced model.
In the action (\ref{action4}), we can integrate the fermionic coordinates $\theta$ and integrate out matrices corresponding to auxiliary fields after an appropriate gauge fixing.
We are them able to obtain the large-$N$ reduced model of the super Yang-Mills theory in the component formalism,
\begin{align}
S=-\frac{1}{g^2} \Tr \left( \frac{1}{4} [\hat{A}_\mu, \hat{A}_\nu ]^2
+ \frac{1}{2} \bar{\hat{\Psi}} \Gamma^\mu[ \hat{A}_\mu, \hat{\Psi}] \right),
\end{align}
where $\mu=1,\cdots,4$, and $\hat{\Psi}$ is the super partner of $\hat{A}_\mu$.
The super Yang-Mills theory on the non-commutative space satisfying $[\hat{x}^\mu, \hat{x}^\nu] =-iC^{\mu\nu}$ can be derived from this model as Appendix \ref{a1}.
However, we can directly derive it in the superfield formalism as follows.\cite{KKM}

First, by the variation of $e^{\hat{\V}}$ in (\ref{action3}),
we obtain the equation of motion 
\begin{align}
\left( \delta e^{\hat{\V}}  \right) e^{-\hat{\V}}  \frac{\partial}{\partial \theta_\alpha}
\left( e^{\hat{\V}} \left( -\frac{1}4   \frac{\partial}{\partial \bar{\theta}^{\dot{\alpha}}} \frac{\partial}{\partial \bar{\theta}_{\dot{\alpha}}} e^{-\hat{\V}} \frac{\partial}{\partial \theta^\alpha} e^{\hat{\V}} \right) e^{-\hat{\V}}  \right)=0. 
\end{align}
This equation has the classical solution
\begin{align}
e^{\hat{\V}}= e^{-2 \theta \sigma^\mu \bar{\theta} \hat{p}_\mu \otimes 1_n }.
\end{align}
This solution is equivalent to $\hat{A}_{\mu} = \hat{p}_{\mu}\otimes 1_n$ (\ref{solution2}) in the component formalism.
If $\hat{\theta}^\alpha$ and $\bar{\hat{\theta}}^{\dot{\alpha}}$ do not anti-commute, this solution does not satisfy the equation of motion (\ref{eom}). 
Therefore, we need the commutative limit.

Note that  this solution implies
\begin{align}
\hat{A}_\alpha = 
e^{ \hat{\theta} \sigma^\mu \bar{\hat{\theta}} \hat{p}_\mu }
\hat{\pi}_\alpha
e^{- \hat{\theta} \sigma^\mu \bar{\hat{\theta}} \hat{p}_\mu }
\end{align}
in the original supermatrix model.

We can expand $e^{\hat{\V}}$ about this solution as
\begin{align}
e^{\hat{\V}}= e^{- \theta \sigma^\mu \bar{\theta} \hat{p}_\mu } e^{\hat{V}} e^{- \theta \sigma^\mu \bar{\theta} \hat{p}_\mu } 
=e^{\hat{A}} e^{\hat{V}} e^{\hat{A}}, 
\end{align}
where we have defined $\hat{A}\equiv - \theta \sigma^\mu \bar{\theta} \hat{p}_\mu$.
Then, the action becomes
\begin{align}
S= &\frac{8\hat{N} \det{\gamma}}{g_m}  \int d^2\theta ~2\pi i \tau\Tr
\left(-\frac{1}4 
\frac{\partial}{\partial \bar{\theta}^{\dot{\alpha}}} 
\frac{\partial}{\partial \bar{\theta}_{\dot{\alpha}}}
e^{ -\hat{A} } 
e^{-\hat{V}}
e^{ -\hat{A}}
 \frac{\partial}{\partial \theta^\alpha}
 e^{\hat{A} } 
 e^{\hat{V}} 
 e^{\hat{A}} 
 \right)^2 +\text{c.c.}\nonumber \\
=&\frac{8\hat{N} \det{\gamma}}{g_m}  \int d^2\theta ~2\pi i \tau\Tr
\left(-\frac{1}4 
\frac{\partial}{\partial \bar{\theta}^{\dot{\alpha}}} 
\frac{\partial}{\partial \bar{\theta}_{\dot{\alpha}}}
e^{ -\hat{A} } 
e^{-\hat{V}}
e^{ \hat{A}}
e^{ -2\hat{A}}
 \frac{\partial}{\partial \theta^\alpha}
 e^{2\hat{A} }
 e^{ -\hat{A}} 
 e^{\hat{V}} 
 e^{\hat{A}} 
 \right)^2 +\text{c.c.}.
\end{align}
Here it is convenient to use the following equations:
\begin{align}
e^{- \hat{A}}
f(\hat{x})e^{ \hat{A}}
&=e^{ \theta \sigma^\mu \bar{\theta} \hat{p}_\mu }
f(\hat{x})
e^{ -\theta \sigma^\mu \bar{\theta} \hat{p}_\mu }
= f(\hat{x} -i \theta\sigma^\mu \bar{\theta}  ), \\
e^{ -2 \hat{A}}
\frac{\partial}{\partial \theta^\alpha}
e^{2\hat{A} }&=\frac{\partial}{\partial \theta^\alpha} - 2 \sigma^\mu_{\alpha \dot{\alpha}}
\bar{\theta}^{\dot{\alpha}} \hat{p}_\mu -2 (\theta \sigma^\mu \bar{\theta}) 
\sigma^\nu_{\alpha \dot{\alpha}} \bar{\theta}^{\dot{\alpha}} i B_{\mu\nu}, \\
\frac{\partial}{\partial \bar{\theta}^{\dot{\alpha}}}
 e^{-\hat{V}} \hat{p}_\mu  e^{\hat{V}}&=
 \frac{\partial}{\partial \bar{\theta}^{\dot{\alpha}}}
  e^{-\hat{V}} [ \hat{p}_\mu , e^{\hat{V}} ] + \frac{\partial}{\partial \bar{\theta}^{\dot{\alpha}}} \hat{p}_\mu = \frac{\partial}{\partial \bar{\theta}^{\dot{\alpha}}}
  e^{-\hat{V}} [ \hat{p}_\mu , e^{\hat{V}} ],
\end{align}
where the matrix $f(\hat{x})$ corresponds to the non-commutative field $f(x)$ via the Weyl ordering (\ref{Weyl2}).
Now, through the mapping rules given in Appendix \ref{a1}, we can map this reduced model to the super Yang-Mills theory on the non-commutative space, whose action is
\begin{align}
S=&\frac{8\hat{N} \det{\gamma}}{g_m (2\pi)^2 \sqrt{\det C} }2\pi i \tau \times\nonumber \\
&  \int d^4x d^2\theta ~\Tr_{U(n)}
\left(-\frac{1}4 
\frac{\partial}{\partial \bar{\theta}^{\dot{\alpha}}} 
\frac{\partial}{\partial \bar{\theta}_{\dot{\alpha}}}
e^{-V(x-i \theta \sigma \bar{\theta},\theta, \bar{\theta})}
\left( \frac{\partial}{\partial \theta^\alpha} + 2i \sigma^\mu_{\alpha \dot{\alpha}}
\bar{\theta}^{\dot{\alpha}} \partial_\mu \right) 
  e^{V(x-i \theta \sigma \bar{\theta},\theta, \bar{\theta})} 
 \right)^2_* \nonumber \\
 & +\text{c.c.}.
\end{align}
Then, it is reasonable to identify this $x^\mu$ as $y^\mu~(= x^\mu + i \theta \sigma^\mu \bar{\theta})$ in the ordinary notation,
since
\begin{align}
D_\alpha = \frac{\partial}{\partial \theta^\alpha} + 2i \sigma^\mu_{\alpha \dot{\alpha}}
\bar{\theta}^{\dot{\alpha}} \frac{\partial}{\partial y^\mu}, ~~\bar{D}_{\dot{\alpha}}= - \frac{\partial}{\partial \bar{\theta}^{\dot{\alpha}}}
\end{align}
in the $y^\mu$ expression.

In addition, it is known that $\sqrt{\det C}$ is proportional to $\hat{N}$\cite{KKM}, and therefore we take the constant $g_m$ as
\begin{align}
\frac{\hat{N}}{g_m}=
\frac{ (2\pi)^2 \sqrt{\det C} }{8 \det{\gamma}}.
\end{align}
Then, we obtain the action of the super Yang-Mills theory,\footnote{
When we study the reduced model, the covariant approach is convenient.
However, we derive the action of the ordinary approach, because this action possesses a simpler expression.}
\begin{align}
S= \int d^4y d^2 \theta~ 2\pi i \tau \Tr \left(-\frac{1}{4}\bar{D}_{\dot{\alpha}} \bar{D}^{\dot{\alpha}} e^{-V} D_\alpha e^V   \right)^2_* +\text{c.c.}.
\label{SYM7}
\end{align}
Thus, we have derived the super Yang-Mills theory from the supermatrix model.

In the above derivation, it was necessary to take the commutative limit $\gamma^{\alpha\dot{\alpha}} \rightarrow 0$.
However, if we consider the super Yang-Mills theory (\ref{SYM7}) on 4-dimensional non-commutative superspace $(x^\mu, \theta, \bar{\theta} )$, it may not be necessary to take this limit.
Actually, as we show in \S \ref{2.4}, field theories on such a superspace can be defined.
We might be able to find a classical solution in the supermatrix model which is such that the action expanded about it can be directly mapped to the super Yang-Mills theory on the non-commutative superspace.
In order to find a proper solution, we can also add appropriate CS terms to the action of the supermatrix model, as done in the studies of fuzzy spheres.\cite{FS}

Finally, we discuss the symmetry of this theory.
The super Yang-Mills theory possesses $\N=1$ supersymmetry and $U(n)$ local gauge symmetry.
The supersymmetry is generated by the derivatives $\partial/\partial x^\mu$, $\partial/\partial \theta^\alpha$ and $\partial/\partial \bar{\theta}^{\dot{\alpha}}$, and the gauge symmetry is generated by a (anti-)chiral superfield which is a function of $(x^\mu, \theta^\alpha, \bar{\theta}^{\dot{\alpha}})$.
Thus, these two symmetries are quite different in ordinary field theory.
However, the coordinates $(x^\mu, \theta^\alpha, \bar{\theta}^{\dot{\alpha}})$ and their momentum are essentially equivalent in the supermatrix model.
Therefore, these two symmetries are unified into the $GL(2|2,R)\times U(\hat{N})$ gauge symmetry (\ref{symmetry}) in the original supermatrix model.
Note that this supersymmetry is different from the original $\N=1$ supersymmetry, 
since we use the non-commutative superspace and the commutation relations between the supercharges are deformed.
When we take the commutative limit $\gamma^{\alpha\dot{\alpha}} \rightarrow 0$, the original supersymmetry is reproduced.

\section{Matter superfields and extended supersymmetries}
\label{2.3}
\setcounter{equation}{0}

In this section, we consider models coupled to matter superfields.
In order to do this, we define supermatrices $\hat{\Phi}$ and $\bar{\hat{\Phi}}$ corresponding to the (anti-)chiral superfields $\Phi$ and $\bar{\Phi}$ in the covariant approach.
We know that the constraint equation of the matter (anti-)chiral superfield belonging to the adjoint representation is $\bar{\nabla}_{\dot{\alpha}}\Phi=0$ ($\nabla_\alpha \bar{\Phi}=0$) (\ref{constraint3}). 
As we studied in \S \ref{2.0}, the supermatrices $\hat{A}_\alpha$ and $\bar{\hat{A}}_{\dot{\alpha}}$ correspond to the covariant derivatives as
\begin{align}
\hat{A}_\alpha \leftrightarrow \nabla_\alpha ,~
\bar{\hat{A}}_{\dot{\alpha}} \leftrightarrow \bar{\nabla}_{\dot{\alpha}}. 
\end{align}
Accordingly, the constraint equations become
\begin{align}
[\bar{\hat{A}}_{\dot{\alpha}}, \hat{\Phi} ] =0,~
[\hat{A}_\alpha, \bar{\hat{\Phi}}  ]=0
\label{constraint4}
\end{align}
in the supermatrix model.

Comparing the action of super Yang-Mills theory (\ref{action5}),
we can obtain the action of the supermatrix model,
\begin{align}
S=\frac{\hat{N}}{g_m}\Str \left( 2\pi i \tau
\{ \hat{A}_\alpha , \bar{\hat{A}}_{\dot{\alpha}} \}^2 
+ 2\bar{\hat{\Phi}} \hat{\Phi} \right)
+ \frac{\hat{N}}{g_m}2 \Str' \left( W(\hat{\Phi})  \right)
+\text{c.c.},
\label{action6}
\end{align}
where $W(\hat{\Phi})$ is a superpotential and 
 $\Str'$ is a new supertrace.
In order to construct the supertrace corresponding to the F-term integral, $\int d^4x d^2 \theta$,
we have to introduce new matrices $\hat{\Theta}^\alpha$ and $\bar{\hat{\Theta}}^{\dot{\alpha}}$ satisfying the constraints
\begin{align}
\{ \hat{A}_\alpha, \hat{\Theta}^\beta \}= \delta_\alpha^\beta,~ 
\{ \bar{\hat{A}}_{\dot{\alpha}}, \bar{\hat{\Theta}}^{\dot{\beta}} \}
=\delta_{\dot{\alpha}}^{\dot{\beta}}.
\end{align}
Accordingly, when we expand $\hat{A}_\alpha$ about the classical solution (\ref{solution5}) as $\hat{A}_\alpha=e^{-\hat{\omega}}\hat{\pi}_\alpha e^{\hat{\omega}}$, 
this constraint requires $\hat{\Theta}^\alpha=e^{-\hat{\omega}}\hat{\theta}^\alpha e^{\hat{\omega}}$.
Using this relation, we define $\Str'$ as
\begin{align}
\Str'\hat{O} \equiv \Str\left( \bar{\hat{\Theta}}^2 \hat{O} \right).
\end{align}
Then, we can show that this supertrace corresponds to the integral as follows.
When we consider the model expanded about the classical solution, 
we can solve the constraint equations (\ref{constraint4}) as $ \hat{\Phi}=e^{\bar{\hat{\omega}}}\hat{\Phi}_0e^{-\bar{\hat{\omega}}}$ and $\bar{\hat{\Phi}}=e^{-\hat{\omega}} \bar{\hat{\Phi}}_0 e^{\hat{\omega}} $, where $\hat{\Phi}_0$ and $\bar{\hat{\Phi}}_0$ satisfy the constraints
\begin{align}
[\bar{\hat{\pi}}_{\dot{\alpha}}, \hat{\Phi}_0]=0,~[\hat{\pi}_\alpha, \bar{\hat{\Phi}}_0]=0.
\label{constraint11}
\end{align}
Then, the supertrace becomes
\begin{align}
\Str'W(\hat{\Phi})=  \Str\left( \bar{\hat{\theta}}^2 W(\hat{\Phi}_0)  \right)
=&4 \det \gamma \int  d^2 \theta d^2 \bar{\theta}~  \Tr_{U(\hat{N})}\left(\bar{\theta}^2  W(\hat{\Phi}_0(\theta)) \right)_\star \nonumber \\
=& 4 \det \gamma \int  d^2 \theta~ \Tr_{U(\hat{N})}\left(  W(\hat{\Phi}_0(\theta)) \right)_\star,
\end{align}
where we have used the mapping rules in \S \ref{3.1} and $\hat{\Phi}_0$ is mapped to $\hat{\Phi}_0(\theta)$.\footnote{The constraint (\ref{constraint11}) implies that $\hat{\Phi}_0(\theta)$ does not depend on $\bar{\theta}$.}
Actually, we can show that the last equation corresponds to the F-term of (\ref{action}) in the commutative limit.
Because (\ref{action}) and (\ref{action5}) are equivalent, the model (\ref{action6}) corresponds to the super Yang-Mills theory (\ref{action5}).

We can also extend the gauge group and couple fundamental and/or bifundamental matter superfields to this model by employing the argument given in Ref. \citen{KKM2}, 
\\

Here we discuss the relation to the Dijkgraaf-Vafa theory\cite{DV}, which asserts the equivalence of the supersymmetric gauge theories and the corresponding matrix models.
By considering our model (\ref{action6}) expanded about the classical solution (\ref{solution5}), we  obtain
\begin{align}
S=& \frac{\hat{N}}{g_m}2\Str \left(e^{-\hat{\V}} \bar{\hat{\Phi}}_0 e^{\hat{\V}} \hat{\Phi}_0 \right) \nonumber \\
&+ \frac{\hat{N}}{g_m} 2\Str~\bar{\hat{\theta}}^2 \left( 
2\pi i \tau \left( 
-\frac{1}4 
[ \bar{\hat{\pi}}_{\dot{\alpha}},
\{ \bar{\hat{\pi}}^{\dot{\alpha}} ,
e^{-\hat{\V}} 
[ \hat{\pi}_{\alpha}
, e^{\hat{\V}} ] \} ]
\right)^2+
W(\hat{\Phi}_0)  \right)
+\text{c.c.}.
\end{align}
This model is similar to the supermatrix model studied in Ref. \citen{KKM}.
That model was introduced to prove the Dijkgraaf-Vafa theory.
These two supermatrix models correspond to field theories on non-commutative superspaces, but the types of non-commutativity are different.
Specifically, our coordinates satisfy (\ref{NCSS4}), whereas theirs satisfy (\ref{NCSS3}).
Accordingly, the gauge groups are also different.
Our group is $GL(2|2,R)\times U(\hat{N})$ and their is $U(1|1)\times U(1|1) \times U(\hat{N})$.
However, the planar diagrams do not depend on the non-commutativity\cite{TM}. 
Therefore, these two models are equivalent with respect to such quantities.
Furthermore, by considering only the planar diagrams in the matrix model, we can prove the Dijkgraaf-Vafa theory.
In this sense, we conjecture that this equivalence can also be shown through our supermatrix model.
However, when we consider the non-planar diagrams that depend on the non-commutativity, these two models are distinguishable.\footnote{It is known that the contribution of the non-planar diagrams in the matrix model corresponds to the contribution of the graviphoton in supersymmetric field theory.\cite{OV2, TM}
Such a correspondence is not clear in our model. }\\

{\bf $ \N =2$ super Yang-Mills theory}

By taking $W(\hat{\Phi})=0$ in (\ref{action6}) and rescaling $\hat{\Phi}$, 
we can obtain the action of a supermatrix model corresponding to $\N=2$ super Yang-Mills theory,
\begin{align}
S=&\frac{\hat{N}}{g_m} 2 \pi i \tau \Str \left(
\{ \hat{A}_\alpha , \bar{\hat{A}}_{\dot{\alpha}} \}^2 
+2 \bar{\hat{\Phi}} \hat{\Phi} \right).
\label{action7}
\end{align}

Let us investigate the $\N=2$ supersymmetry.
As mentioned in \S \ref{2.0}, the $GL(2|2,R)\times U(\hat{N})$ symmetry contains the deformed $\N=1$ supersymmetry.
This action is also invariant under the additional four fermionic transformations
\begin{align}
&\delta_{\epsilon} \hat{A}_\alpha = \epsilon_{\alpha} \bar{\hat{\Phi}}, \nonumber \\ 
&\delta_{\epsilon} \hat{\Phi} = - \epsilon^{\alpha} [ \bar{\hat{A}}_{\dot{\alpha}} , \{ \bar{\hat{A}}^{\dot{\alpha}} , \hat{A}_\alpha \}   ], \nonumber \\
&\delta_{\bar{\epsilon}} \bar{\hat{A}}_{\dot{\alpha}} = \bar{\epsilon}_{\dot{\alpha}} \hat{\Phi}, \nonumber \\
&\delta_{\bar{\epsilon}} \bar{\hat{\Phi}} = - \bar{\epsilon}_{\dot{\alpha}}
[ \hat{A}^\alpha, \{ \hat{A}_\alpha, \bar{\hat{A}}^{\dot{\alpha}}   \} ].
\label{N=2}
\end{align}
It can also be confirmed that the constraints (\ref{constraint2}) and (\ref{constraint4}) are invariant by using Jacobi identities.
Iterating these transformations, we obtain
\begin{align}
\left(\delta_{\bar{\epsilon}} \delta_{\epsilon} -\delta_{\epsilon}\delta_{\bar{\epsilon}} \right) \hat{\Phi} = 2 \epsilon^\alpha \bar{\epsilon}_{\dot{\alpha}}
[\{ \bar{\hat{A}}^{\dot{\alpha}}, \hat{A}_\alpha  \} , \hat{\Phi} ].
\end{align}
Because $\hat{A}_{\alpha}$ corresponds to $\nabla_\alpha$, the anti-commutator $\{ \bar{\hat{A}}_{\dot{\alpha}}, \hat{A}_\alpha  \} $ should be regarded as the generator of the translation.
Thus, this equation implies that the fermionic symmetries represent the supersymmetry.
Therefore, this model possesses the deformed $\N=2$ supersymmetry.\\

{\bf $\N=4$ super Yang-Mills theory}

The action of the supermatrix model corresponding to $\N=4$ super Yang-Mills theory is given by
\begin{align}
S=&\frac{\hat{N}}{g_m}  \Str \left(
\{ \hat{A}_\alpha , \bar{\hat{A}}_{\dot{\alpha}} \}^2 
+ 2\bar{\hat{\Phi}}_1 \hat{\Phi}_1 + 2\bar{\hat{\Phi}}_2 \hat{\Phi}_2+ 2\bar{\hat{\Phi}}_3 \hat{\Phi}_3 \right) \nonumber \\
&+ \frac{\hat{N}}{g_m} 2\Str' \left( \hat{\Phi}_1 [ \hat{\Phi}_2, \hat{\Phi}_3  ]  \right) +\text{c.c.},
\label{N=4}
\end{align}
where the matrices $\hat{\Phi}_i$ are the adjoint matter supermatrices satisfying the constraint $[\bar{\hat{A}}_{\dot{\alpha}},\hat{\Phi}_i ]$.

It is known that the IIB matrix model\cite{IKKT} is obtained from $\N=4$ $U(\hat{N})$ super Yang-Mills theory through the large-$N$ reduction.
Therefore, we conjecture that our model is equivalent to the IIB matrix model.

It has been shown that the $\N=4$ $U(\hat{N})$ super Yang-Mills theory also has a close relation to the matrix models in studies of the AdS/CFT correspondence.\cite{half}
It might be possible to obtain an understanding of this relation through our model.

\section{Discussion}
\setcounter{equation}{0}
\label{Discussions}

We showed that super Yang-Mills theories in the superfield formalism can be derived from the supermatrix models.
In this derivation, a new non-anti-commutative superspace is generated from the supermatrix model.
We investigate this space in \S \ref{2.4}.
In \S \ref{Applications}, we consider applications of our study.

\subsection{Construction of field theory on the non-commutative superspace}
\setcounter{equation}{0}
\label{2.4}
Here we discuss two points concerning of the non-anti-commutative superspace.
The first is the reason that we consider the non-commutative superspace (\ref{NCSS2}).
The second is the way to define field theories on 4-dimensional non-commutative superspace.\\

First we discuss the question of why we choose 
\begin{align}
\{ \hat{\theta}^{\alpha} , \bar{\hat{\theta}}^{\dot{\alpha}} \} = \gamma^{\alpha \dot{\alpha}},~ 
\{ \hat{\theta}^\alpha , \hat{\theta}^\beta \}= \{ \bar{\hat{\theta}}^{\dot{\alpha}} , \bar{\hat{\theta}}^{\dot{\beta}} \}=0,
\label{NCSS4}
\end{align}
instead of the ordinary non-commutative superspace $\{\theta^\alpha , \theta^\beta \} \ne 0 $ \cite{S, KKM} in this study.
As in Ref. \citen{KKM}, let us suppose that the non-commutative coordinates satisfy the anti-commutation relations
\begin{align}
\{\hat{ \theta}^\alpha , \hat{\theta}^\beta \} = \gamma^{\alpha \beta},~
 \{ \bar{\hat{\theta}}^{\dot{\alpha}} , \bar{\hat{\theta}}^{\dot{\beta}} \}=
 \gamma^{* \dot{\alpha} \dot{\beta}}, ~ 
 \{\hat{ \theta}^{\alpha} , \bar{\hat{\theta}}^{\dot{\alpha}} \} = 0.
 \label{NCSS3}
\end{align}
Then, we can define matrices $ \hat{\pi}_\alpha$ and $\bar{\hat{\pi}}_{\dot{\alpha}}$ as
\begin{align}
\hat{\pi}_{\alpha} = \beta_{\alpha \beta} \hat{\theta}^\beta,~
\bar{\hat{\pi}}_{\dot{\alpha}} = \beta^*_{\dot{\alpha} \dot{\beta}} \bar{\hat{\theta}}^{\dot{\beta}},
\end{align}
corresponding to the derivative operators, where $\beta_{\alpha \beta}$ is the inverse matrix of $\gamma^{\alpha \beta}$:
\begin{align}
\beta_{\alpha\beta}\gamma^{\beta\gamma}=\delta_\alpha^{\gamma},~
\beta^*_{\dot{\alpha}\dot{\beta}} \gamma^{\dot{\beta} \dot{\gamma}}=\delta_{\dot{\alpha}}^{\dot{\gamma}}.
\end{align}
Then, these matrices satisfy $\{ \hat{\pi}_\alpha, \hat{\theta}^{\beta} \} = \delta_\alpha^\beta$ and $\{  \hat{\pi}_\alpha,  \hat{\pi}_\beta \} = \beta_{\alpha \beta} $.
As in \S \ref{2.2}, we conjecture that $\hat{A}_{\alpha}=  \hat{\pi}_\alpha \otimes 1_{\hat{N}} $ is a classical solution of the model.
However, because we have $\{  \hat{\pi}_\alpha,  \hat{\pi}_\beta \} = \beta_{\alpha \beta} $, this does not satisfy the constraint $\{ \hat{A}_{\alpha}, \hat{A}_{\beta} \}=0$.
Consequently, we cannot use the same procedure.
This is the reason that we chose the non-commutative superspace (\ref{NCSS4}).\footnote{I would like to thank H. Kawai for valuable discussions about the non-commutative superspace.}\\

Now we consider the definition of the 4-dimensional field theories on non-commutative superspace.
In \S \ref{2.0}, we studied the reduced model on non-commutative superspace, but we did not consider field theories on this space.
However, in deriving the super Yang-Mills theory, if we could avoid taking the commutative limit $\gamma^{\alpha \dot{\alpha}} \rightarrow 0$, we would obtain it.
With this in mind, we discuss the definition of field theories on this space.

Among similar studies of such a definition, Seiberg's work\cite{S} is well known. 
He investigated the non-commutative superspace (\ref{NCSS3}) with $\gamma^{* \dot{\alpha} \dot{\beta}} =0$.
In addition, he imposed the condition $[x^\mu, \theta]\ne 0 $ so that $y^\mu$ and $\theta$ satisfy $[y^\mu, \theta ]=0$.
Then, he showed that the chiral and anti-chiral superfields can be defined and that field theories on such a space can be constructed consistently.
However, this construction cannot be applied to the case of non-zero $\gamma^{* \dot{\alpha} \dot{\beta}}$ or our superspace (\ref{NCSS4}).
We show that we can construct field theories even on these spaces by introducing non-local fields.

First, we reconsider the (anti-)chiral superfield on the ordinary commutative superspace.\footnote{In this subsection, we do not use the covariant approach but investigate the ordinary approach.}
These fields satisfy the constraints
\begin{align}
\bar{D}_{\dot{\alpha}} \Phi=0,~D_\alpha \bar{\Phi}=0.
\end{align}
Here the derivatives $D_\alpha$ and $\bar{D}_{\dot{\alpha}}$ can be rewritten as\cite{KKM} 
\begin{align}
D_\alpha &= \frac{\partial}{\partial \theta^\alpha} + i \sigma^\mu_{\alpha \dot{\alpha} } \bar{\theta}^{\dot{\alpha}} \partial_\mu= e^{-A}   \frac{\partial}{\partial \theta^\alpha} e^{A}, \nonumber \\
\bar{D}_{\dot{\alpha}} &= - \frac{\partial}{\partial \bar{\theta}^{\dot{\alpha}} }
-i \theta^\alpha \sigma^\mu_{\alpha \dot{\alpha} } \partial_\mu = e^{A} \left(- \frac{\partial}{\partial \bar{\theta}^{\dot{\alpha}} } \right) e^{-A},
\label{derivative}
\end{align}
where $A\equiv i \theta \sigma^\mu \bar{\theta} \partial_\mu$.
Then, the constraint equations are solved as
\begin{align}
\Phi= \Phi( e^A x e^{-A}, e^{A} \theta e^{-A}  ),~~\bar{\Phi}=\bar{\Phi}( e^{-A} x e^{A}, e^{-A} \bar{\theta} e^{A}  ).
\label{solution6}
\end{align}
Here, we can expand the exponential as
\begin{align} 
e^A x^\mu e^{-A}&= x^\mu +   i \theta \sigma^\mu \bar{\theta}= y^\mu,~
e^A \theta^\alpha e^{-A} = \theta^\alpha , \nonumber \\
e^{-A} x^\mu e^{A}&= x^\mu -   i \theta \sigma^\mu \bar{\theta}= \bar{y}^\mu,~
e^{-A} \bar{\theta}^{\dot{\alpha}} e^{A} = \bar{\theta}^{\dot{\alpha}},
\end{align}
and (\ref{solution6}) becomes the well-known solution. 
Next we apply it to the definition of the (anti-)chiral superfield on the non-commutative superspace (\ref{NCSS4}).
We use the right-hand side of (\ref{derivative}) to define $D$ and $\bar{D}$ on the non-commutative superspace as
\begin{align}
D_\alpha & \equiv   e^{-A} \star  \frac{\partial}{\partial \theta^\alpha} \star e^{A}= \frac{\partial}{\partial \theta^\alpha} + i \sigma^\mu_{\alpha \dot{\alpha} } \bar{\theta}^{\dot{\alpha}} \partial_\mu +\cdots,  \nonumber \\
\bar{D}_{\dot{\alpha}} & \equiv e^{A} \star \left(- \frac{\partial}{\partial \bar{\theta}^{\dot{\alpha}} } \right) \star e^{-A}  = - \frac{\partial}{\partial \bar{\theta}^{\dot{\alpha}} }
-i \theta^\alpha \sigma^\mu_{\alpha \dot{\alpha} } \partial_\mu +\cdots ,
\end{align}
where we have used the $\star$-product (\ref{star}).
Here we need to define $A$ as the Weyl ordered form (\ref{Weyl}), since $A$ is a function of $\theta$ and $\bar{\theta}$.
The `$\cdots$' in the right-hand sides of these equations represent series containing differential operators and the non-commutative parameter.
Now we can define the (anti-)chiral superfield so as to satisfy the constraint $\bar{D}_{\dot{\alpha}} \star \Phi=0$ ($D_\alpha \star \bar{\Phi}=0$).
Then, these constraint equations can be solved as
\begin{align}
\Phi&= \Phi( e^A \star x \star e^{-A}, e^{A}\star \theta \star e^{-A}  ) 
=e^A \star \Phi(x,\theta)\star e^{-A} ,\nonumber \\
\bar{\Phi}&=\bar{\Phi}( e^{-A} \star x \star e^{A}, e^{-A} \star \bar{\theta} \star e^{A}  )
=e^{-A} \star \bar{\Phi}(x,\bar{\theta}) \star e^A .
\label{chiral}
\end{align}
Here, $ e^A \star x \star e^{-A} $,  $ e^{A}\star \theta \star e^{-A} $,
$ e^{-A} \star x \star e^{A}$ and  $e^{-A} \star \bar{\theta} \star e^{A}$ can be expanded in terms of the coordinates and the differential operators.
Accordingly, the (anti-)chiral superfield is defined as the non-local field.
With this definition, the (anti-)chiral superfield possesses the chiral ring property.
In other words, a product of chiral superfields is also a chiral superfield.
In this sense, this definition is consistent.
However, evidently this theory does not possess the ordinary supersymmetry.

If we wish to consider field theories on another non-commutative superspace, (\ref{NCSS3}), we have to define the Weyl ordering and the new $\star$-product as \cite{KKM}
\begin{align}
O(\theta, \bar{\theta})&= \int 2^4 d^2 \kappa d^2\bar{\kappa} ~ \tilde{O}(\kappa, \bar{\kappa}) 
e^{\kappa^\alpha \theta_\alpha + \bar{\kappa}_{\dot{\alpha}} \bar{\theta}^{\dot{\alpha}} },\\
O_1\star O_2(\theta , \bar{\theta} )&=
 \left. \exp\left( -\frac{1}{2} \gamma^{\alpha \beta }  \frac{\partial}{\partial \theta_1^\alpha}
  \frac{\partial}{\partial \theta_2^\beta  }
   -\frac{1}{2} \gamma^{\dot{\alpha} \dot{\beta} }   \frac{\partial}{\partial \bar{\theta}_1^{ \dot{\alpha}}}
  \frac{\partial}{\partial \bar{\theta}_2^{\dot{\beta}}  } \right)  
   O_1(\theta_1 , \bar{\theta}_1 )O_2(\theta_2 , \bar{\theta}_2 ) \right|_{\theta=\theta_1=\theta_2 }. 
\end{align}
Using these, we can also define the chiral superfield on this superspace.

Next, in order to consider gauge theory, we define a vector superfield.
We can simply define it by $\bar{V}=V$ and the Weyl ordered form. 
Then, we can construct the gauge invariant action
\begin{align}
S=&\int d^4x d^2 \theta d^2 \bar{\theta}~ 
\Tr \left( e^{-A} \bar{\Phi}(x, \bar{\theta})e^A  e^{V(x,\theta, \bar{\theta})} e^A \Phi(x,\theta) e^{-A} e^{-V(x,\theta, \bar{\theta})}\right)_\star \nonumber \\
&+\int  d^4x d^2 \theta~ \Tr \left( 2\pi i \tau  W^\alpha W_\alpha + W(\Phi) \right)_\star+\text{c.c.}.
\end{align}
Consequently, field theories on the 4-dimensional non-commutative superspace (\ref{NCSS4}) and (\ref{NCSS3}) can be defined consistently by using the above non-local chiral superfield.
In addition, we can study a deformation satisfying $[x^\mu, x^\nu]\ne 0$ using the ordinary procedure.

We have shown that we can define gauge theory on the 4-dimensional non-commutative superspace.
As stated in \S \ref{2.0}, we have not yet derived it from the supermatrix model.
For further study of this derivation, this definition would be important.

\subsection{Applications}
\label{Applications}

In this paper, we have constructed new supermatrix models.
These models are regarded as the large-$N$ reduction of the super Yang-Mills theory in the superfield formalism.
We should also be able to construct such supermatrix models in other supersymmetric theories.
It is known that the large-$N$ reduced models are important in studies of non-perturbative aspects of gauge theory and string theory.
Thus, we believe that such supermatrix models are also important in supersymmetric theories.
For example, these models may help in the construction of supersymmetric lattice gauge theories.
It is known that lattice gauge theory can be obtained from the reduced model by the orbifolding.\cite{Kaplan} \ 
It would be interesting to apply it to our models.

The application to Dijkgraaf-Vafa theory \cite{KKM, DV, KKM2} is also interesting.
In this theory, the free energy of the matrix models corresponds to the pre-potential of supersymmetric gauge theories.
We can derive the low energy effective action of these gauge theories from the pre-potential.
In order to evaluate the free energy precisely, we need to define the measure of the path integral.
However, this definition is not clearly understood.
For this reason, we need to define it by hand such that it reproduces the known effective theories, as in $\N=1$\cite{DV, OV3} or $\N=4$ super Yang-Mills theory.\cite{KKMY} \ 
In this sense, we cannot derive all the information regarding gauge theories from the matrix models.
However, super Yang-Mills theories can be directly derived from our supermatrix models.
Therefore, we can evaluate the measure in our models.
In addition, our model may allow us to understand why the free energy corresponds to the pre-potential.\\

In our work, we can add CS terms to the action of the supermatrix models.
Then we would find classical solutions corresponding to the configuration of the fuzzy superspheres.\cite{NCSS} \ 
Such a study would be interesting in the investigation of supermanifolds.\\

Finally, we note that one interesting operator of this model is the Wilson loop,
\begin{align}
\Str \left( \prod_{i}\exp \left( \epsilon_i^\alpha \hat{A}_{\alpha} + \epsilon_i^{\dot{\alpha}}
\bar{\hat{A}}_{\dot{\alpha}} \right) \right).
\end{align}
The Wilson loop is an important operator in gauge theory, and for this reason, we believe that it is also important in our model.

\section*{Acknowledgements}
I would like to thank my colleagues in the Theoretical Particle Physics Group
at Kyoto University and in the Theory Group at KEK for discussions and encouragement.
I am especially grateful to H. Kawai for various useful discussions.
I am also thankful to T. Higaki, S. Iso, T. Kuroki, Y. Matsuo, A. Miwa and H. Shimada for various useful discussions and encouragement. 
This work was supported in part by a JSPS Research Fellowship for Young Scientists and a Grant-in-Aid for the 21st Century COE ``Center for Diversity and Universality in Physics'', Kyoto University.

\appendix

\section{Non-commutative Yang-Mills Theory from the Large-$N$ Reduced Model}
\setcounter{equation}{0}
\label{a1}
In this appendix, we briefly review a derivation of $U(n)$ Yang-Mills theory on non-commutative space from a large-$N$ reduced model.\cite{NCYM} \
Here we consider the non-commutative space satisfying
\begin{align}
[\hat{x}^\mu , \hat{x}^\nu ]= -iC^{\mu\nu},
\end{align}
where $C^{\mu\nu}$  are c-numbers.
In order to construct a derivative operator on this space,
we define the matrix $B_{\mu\nu}$, which is the inverse of $C^{\mu\nu}$.
Then, we can construct the derivative operator from $\hat{x}^\mu$ as
\begin{align}
\hat{p}_\mu= B_{\mu\nu}\hat{x}^\nu.
\end{align}
These matrices satisfy the following commutation relations:
\begin{align}
[\hat{p}_\mu,\hat{x}^\nu]=-i \delta_{\mu}^\nu,~~[\hat{p}_\mu , \hat{p}_\nu ]= iB_{\mu\nu}.
\end{align}
Note that only infinite rank matrices can satisfy these relations.

Now we can derive a $U(n)$ Yang-Mills theory on this space from a $U(\hat{N})$ matrix model with the action
\begin{align}
S=\frac{1}{g^2} \Tr \left(\frac{1}4 [\hat{A}_\mu, \hat{A}_\nu]^2   \right),
\end{align}
where $\hat{A}_\nu$ ($\mu=1\ldots d$) are hermitian matrices.
We take $\hat{N}$ to be infinite and $n$ to be finite and define $\hat{M}$ as $\hat{N} = n \hat{M}$.
It is known that this matrix model is a large-$N$ reduced model of a $d$-dimensional Yang-Mills theory.

This model has the classical solution
\begin{align}
A_\mu=\hat{p}_\mu \otimes 1_n,
\label{solution2}
\end{align}
where $\hat{p}_\mu$ are the $\hat{M} \times \hat{M}$ hermitian matrices defined above.
Then, we define $\hat{a}_\mu$ as fluctuations of $\hat{A}_\mu$ expanded about this solution and obtain the action
\begin{align}
S=\frac{1}{g^2} \Tr \left(\frac{1}4 [\hat{p}_\mu+\hat{a}_\mu, \hat{p}_\nu+\hat{a}_\nu]^2 \right).
\label{fluctuation}
\end{align}
We can map this action to that of a $U(n)$ Yang-Mills theory on the non-commutative space as follows.
First, a matrix $\hat{O}$ can be mapped to the corresponding field $O(x)$ on the non-commutative space via the Weyl ordering:
\begin{align}
\hat{O}=\int \frac{d^d k}{(2\pi)^d} e^{ik_\mu \hat{x}^\mu} \tilde{O}(k)\leftrightarrow
O(x)=\int \frac{d^d k}{(2\pi)^d} e^{ik_\mu x^\mu} \tilde{O}(k).
\label{Weyl2}
\end{align}
Then, a product of two matrices $\hat{O}_1 \hat{O}_2$ corresponds to a $*$-product of two fields:
\begin{align}
\hat{O}_1 \hat{O}_2 \leftrightarrow
O_1 * O_2(x) \equiv
\left. \exp \left( -\frac{i}2 C^{\mu\nu} \frac{\partial}{\partial x^\mu}\frac{\partial}{\partial y^\mu} \right) O_1(x)O_2(y) \right|_{y = x}.
\label{star*}
\end{align}
Also, a commutator $[\hat{p}_\mu,\hat{O}]$ corresponds to $-i\partial_\mu O(x)$.
Finally, a trace is mapped to an integral:
\begin{align}
(2\pi)^{\frac{d}2}\sqrt{\det C}\Tr_{U(\hat{N})}( \hat{O} )&=\int d^4x~\tr_{U(n)}O(x).
\label{int1}
\end{align}
Applying these relations to (\ref{fluctuation}), we can obtain the action of the Yang-Mills theory,
\begin{align}
S=  \int d^{d} x \frac{1}{{\tilde{g}}^2}  \tr_{U(n)} \left(-\frac{1}4 F_{\mu\nu}^2 \right)_*,
\end{align}
where $F_{\mu\nu}$ is the field strength, $*$ indicates that the product of this action is a $*$-product and we define
\begin{align}
\frac{1}{{\tilde{g}}^2} \equiv 
\frac{1}{(2\pi)^{\frac{d}{2}} \sqrt{\det C}} \frac{1}{g^2}.
\end{align}
Note that we ignore a constant term in this action.

The original matrix model has $U(\hat{N})$ gauge symmetry, and thus we have
\begin{align}
\hat{A}'_\mu = e^{i\hat{\Lambda}} \hat{A}_\mu e^{-i \hat{\Lambda}}.
\label{U(N)}
\end{align}
This transformation becomes a local $U(n)$ gauge transformation and translations in the non-commutative Yang-Mills theory.
For example, translations are generated by $\hat{\Lambda}= \epsilon^\mu \hat{p}_\mu $, where the quantities $\epsilon^\mu$ are constants.
Therefore, gauge transformations and translations are unified in the reduced model.

\section{Covariant Approach of $\N=1$ Super Yang-Mills Theory}
\setcounter{equation}{0}
\label{a2}
In this appendix, we review one construction, called the ``covariant approach", of $\N=1$ super Yang-Mills theory.\cite{superspace} \ 
First, we review the conventional approach for constructing it.
Next, we introduce the covariant approach.
Finally, we discuss the relation between these two approaches.

\subsection{Conventional approach}
The action of the super Yang-Mills theory coupled to an adjoint matter superfield $\Phi_0$ in the superfield formalism is described by
\begin{align}
S=\int d^4xd^2\theta d^2 \bar{\theta}~  \Tr \left( \bar{\Phi}_0 e^V \Phi_0 e^{-V}\right) + \int  d^4xd^2\theta ~ \Tr \left( 2\pi i\tau W_0^\alpha W_{0\alpha}+W(\Phi_0) \right) +\text{c.c.},
\label{action}
\end{align}
where $\tau$ is the gauge coupling constant,
$V$ is a vector superfield satisfying the constraint equation $\bar{V}=V$, and $W_{0\alpha}$ is the field strength defined by
\begin{align}
W_{0\alpha}=-\frac{1}{4}\bar{D}^2 e^{-V} D_\alpha e^V.
\label{fieldstrength1}  
\end{align}
Here, $\Phi_0$ and $W_{0\alpha}$ are chiral superfields satisfying the constraints $\bar{D}_{\dot{\alpha}} \Phi_0 =0 $ and $\bar{D}_{\dot{\alpha}} W_{0\alpha}=0$.
When we consider the covariant approach, we define new chiral superfields satisfying new constraints.
We need to distinguish these two chiral superfields.
We use $\Phi_0$ and $W_{0\alpha}$ in the conventional approach and $\Phi$ and $W_\alpha$ in the covariant approach. 

{}From this point, we consider gauge symmetries and covariant derivatives.
This theory remains invariant under local $U(n)$ gauge transformations,
\begin{align}
\Phi'_0 &= e^{-i \Lambda } \Phi_0 e^{i \Lambda}, \nonumber \\
\bar{\Phi}'_0 &= e^{-i \bar{\Lambda} } \bar{\Phi}_0 e^{i \bar{ \Lambda}}, \nonumber \\
 e^{V'} &= e^{-i \bar{\Lambda} } e^V e^{i \Lambda}, 
\label{gauge1}
\end{align}
where $\Lambda$ ($\bar{\Lambda}$) is a (anti-)chiral superfield.
Then, we can define {\it gauge chiral representation covariant derivatives} $\nabla_{cA}$ ($A=(\alpha, \dot{\alpha}, \alpha \dot{\alpha})$) as
\begin{align}
\nabla_{c\alpha} &= e^{-V}D_\alpha e^V \left(=D_\alpha + e^{-V} (D_\alpha e^V ) \right), \\
\bar{\nabla}_{c\dot{\alpha}} &= \bar{D}_{\dot{\alpha}},\\
\nabla_{c\alpha \dot{\alpha}}&= -i \{ \nabla_{c\alpha}, \bar{\nabla}_{c\dot{\alpha}} \}.
\label{covariant1}
\end{align}
They transform under the gauge transformations as
\begin{align}
\nabla'_{c\alpha} &=  e^{-i \Lambda} e^{-V} e^{i \bar{\Lambda} } D_\alpha  e^{-i \bar{\Lambda} } e^V  e^{i \Lambda}
=  e^{-i \Lambda} e^{-V} D_\alpha e^V  e^{i \Lambda} =  e^{-i \Lambda} \nabla_{c\alpha} e^{i \Lambda}, \nonumber \\
\bar{\nabla}'_{c\dot{\alpha}} &=  \bar{D}_{\dot{\alpha}}
= e^{-i \Lambda}  \bar{D}_{\dot{\alpha}} e^{i \Lambda}
=e^{-i \Lambda}  \bar{\nabla}_{c\dot{\alpha}} e^{i \Lambda}, \nonumber \\
\nabla'_{c\alpha \dot{\alpha}}&= -i \{ \nabla'_{c\alpha}, \bar{\nabla}'_{c\dot{\alpha}} \} =e^{-i \Lambda} \nabla_{c\alpha \dot{\alpha}}e^{i \Lambda},
\end{align}
Hence, when these operators act on a chiral superfield $O_0$, 
the resultant superfields $\nabla_{cA} O_0$ transform as chiral superfields. 
For this reason, the operators $\nabla_{cA}$ are called the gauge chiral representation covariant derivatives.
However, when these operators act on an anti-chiral superfield $\bar{O}_0$, the resultant $\nabla_{cA} \bar{O}_0$ do not transform covariantly.
Thus, we need to define the gauge anti-chiral representation covariant derivatives
\begin{align}
\nabla_{a\alpha} = D_\alpha, ~
\bar{\nabla}_{a\dot{\alpha}} = e^V \bar{D}_{\dot{\alpha}} e^{-V},~
\nabla_{a\alpha \dot{\alpha} }= -i \{ \nabla_{a\alpha}, \bar{\nabla}_{a\dot{\alpha}} \}.
\label{covariant2}
\end{align}
These operators transform as $\nabla'_{aA} = e^{-i \bar{\Lambda}} \nabla_{aA} e^{i \bar{\Lambda}} $,
and hence the superfields $\nabla_{aA} \bar{O}_0$ transform covariantly.

Now, the field strength $W_{0\alpha}$ can be described by (anti-)commutators of the covariant derivatives,
\begin{align}
W_{0\alpha} = -\frac{1}{4} [\bar{\nabla}_{c\dot{\alpha}}, \{ \bar{\nabla}_c^{\dot{\alpha}}, \nabla_{c\alpha} \} ], \nonumber \\
\bar{W}_{0\dot{\alpha}} = -\frac{1}{4} [{\nabla}_a^{\alpha}, 
\{ {\nabla}_{a\alpha}, \bar{\nabla}_{a\dot{\alpha}} \} ].
\label{fieldstrength3}
\end{align}

\subsection{Covariant approach}
First, we introduce new gauge covariant derivatives satisfying
\begin{align}
\overline{(\nabla_{\alpha \dot{\alpha}} )} =- \nabla_{\alpha \dot{\alpha}},~
\overline{(\nabla_{\alpha})} = \nabla_{\dot{\alpha}}.
\label{real}
\end{align}
These derivatives transform covariantly under the {\it new} gauge transformations
\begin{align}
\nabla'_A=e^{-iK} \nabla_A e^{iK},
\label{gauge2}
\end{align}
where $K$ is a vector superfield such that the operators $\nabla'_A$ satisfy the conditions (\ref{real}).

Next, we define the connection $\Gamma_A$ and field strength $F_{AB}$ of the operators $\nabla_A$ as 
\begin{align}
\nabla_A&=D_A-i \Gamma_A, \nonumber \\ 
[ \nabla_A, \nabla_B \} &=T_{AB}^C \nabla_C -i F_{AB},
\label{connection}
\end{align}
where the torsion has been defined as
\begin{align}
T_{AB}^C &= 2i\sigma_{\alpha \dot{\alpha}}^\mu, \nonumber \\
&=0 ~~~\text{otherwise} .
\label{torsion}
\end{align}
This torsion is the same as that in the original derivatives, $D_{\alpha \dot{\alpha}}=-2\sigma^\mu_{\alpha \dot{\alpha}} D_\mu =- 2\sigma^\mu_{\alpha \dot{\alpha}} \partial_\mu $, $D_\alpha$ and $\bar{D}_{\dot{\alpha}}$.

Here we need to impose constraints on the covariant derivatives in order to construct a consistent theory.
Next, we solve these constraint equations and obtain  proper connections and field strength.\\
{\bf Conventional constraints}

When we add covariant terms to the covariant derivatives,
the gauge transformations of the covariant derivatives do not change.
Therefore, we can take $F_{\alpha \dot{\alpha}}=0$ by replacing $\Gamma'_{\alpha \dot{\alpha}} = \Gamma_{\alpha \dot{\alpha}} - iF_{\alpha \dot{\alpha}} $ in (\ref{connection}).
This implies
\begin{align}
\nabla_A=(\nabla_\alpha, \bar{\nabla}_{\dot{\alpha}}, -i \{ \nabla_\alpha , \nabla_{\dot{\alpha}} \}). 
\end{align}\\
{\bf Representation-preserving constraints}

Here we consider the definition of (anti-)chiral superfields used in this theory.
We have constructed covariant derivatives, and it is natural to define them covariantly as
\begin{align}
\bar{\nabla}_{\dot{\alpha}} \Phi =0,~
\nabla_{\alpha} \bar{\Phi}=0.
\label{constraint3}
\end{align}
In order for the relation $\bar{\nabla}'_{\dot{\alpha}} \Phi' =0$ to hold, we stipulate that these fields transform under the gauge transformations as
\begin{align}
\Phi' =e^{-iK}\Phi e^{iK} ,~~\bar{\Phi}'= e^{-iK} \bar{\Phi}e^{iK}.
\end{align}
Then, the constraints (\ref{constraint3}) imply
\begin{align}
-i \{ \bar{\nabla}_{\dot{\alpha}}, \bar{\nabla}_{\dot{\beta}} \} \Phi =
F_{\dot{\alpha} \dot{\beta}} \Phi= 0,~ F_{\alpha \beta} \bar{\Phi}= 0.
\end{align}
These equations require the representation-preserving constraints
\begin{align}
F_{\dot{\alpha} \dot{\beta}} &= -i \{ \bar{\nabla}_{\dot{\alpha}}, \bar{\nabla}_{\dot{\beta}} \} =0 , \nonumber \\
 F_{ \alpha \beta}&= -i \{ \nabla_{\alpha}, \nabla_{\beta} \} =0.
\label{constraint1}
\end{align}

These constraint equations can be solved.
The most general solutions are
\begin{align}
\nabla_\alpha&=e^{-\Omega} D_\alpha e^{\Omega} 
\left(=D_\alpha + e^{-\Omega} (D_\alpha e^{\Omega} ) \right)
, \nonumber  \\
\bar{\nabla}_{\dot{\alpha}}&=e^{\bar{\Omega}} \bar{D}_{\dot{\alpha}} e^{-\bar{\Omega}},
\label{covariant}
\end{align}
where $\Omega$ is an arbitrary complex superfield.
Actually, this satisfies
\begin{align}
\{ e^{-\Omega} D_\alpha e^{\Omega} , e^{-\Omega} D_\beta e^{\Omega} \}=
e^{-\Omega}\{ D_\alpha , D_\beta \} e^{\Omega}=0.
\end{align}

We have now obtained a sufficient number of constraints and hence can calculate the Jacobi identities for the covariant derivatives.
Doing so, we obtain the field strengths
\begin{align}
W_\alpha = -\frac{1}{4} [\bar{\nabla}_{\dot{\alpha}}, \{ \bar{\nabla}^{\dot{\alpha}}, \nabla_\alpha \} ] \nonumber, \\
\bar{W}_{\dot{\alpha}}= -\frac{1}{4} [ \nabla^{\alpha}, \{ \nabla_{\alpha}, \bar{\nabla}_{\dot{\alpha}}   \} ],
\label{fieldstrength2}
\end{align}
and also obtain constraints on the (anti-)chiral superfields and the Bianchi identity:
\begin{align}
\bar{\nabla}_{\dot{\alpha}} W_\alpha = \nabla_{\alpha} \bar{W}_{\dot{\alpha}}= 0,\nonumber \\
\nabla^{\alpha} W_\alpha = \bar{\nabla}_{\dot{\alpha}} \bar{W}^{\dot{\alpha}}.
\end{align}

With the above results, we can construct a gauge invariant action for this theory,
\begin{align}
S=\int  d^4xd^2\theta d^2 \bar{\theta}~ \Tr \bar{\Phi}  \Phi  + \int  d^4xd^2\theta~  \Tr \left( 2\pi i \tau W^\alpha W_{\alpha}+W(\Phi) \right)+\text{c.c.}.
\label{action5}
\end{align}
It may seem that in this action, the matter fields are not coupled to the gauge fields. 
However, these fields are coupled to the gauge fields through the constraints (\ref{constraint3}).

Finally, we discuss the gauge symmetry.
The gauge transformation of the covariant derivatives (\ref{gauge2}) requires a transformation of $\Omega$ as
\begin{align}
e^{\Omega'} = e^{\Omega} e^{i K},~
e^{\bar{\Omega}'}= e^{-iK} e^{\bar{\Omega}}.
\label{gauge5}
\end{align}
Also, this theory has an additional gauge symmetry.
The equations given in (\ref{covariant}) are invariant under the transformations
\begin{align}
{e^{\Omega}}'=e^{-i\bar{\Lambda }} e^{\Omega},
~~{e^{\bar{\Omega}}}'= e^{\bar{\Omega}}e^{i\Lambda}, 
\label{gauge3} 
\end{align}
where $\Lambda$ satisfies $\bar{D}_{\dot{\alpha}} \Lambda=0$.
We show below that these transformations become the gauge transformations in the conventional approach, appearing in (\ref{gauge1}).

\subsection{Conventional approach vs. covariant approach}
In this subsection, we study the relations between the conventional approach and the covariant approach.

First, we can $\overline{(e^{\Omega} e^{\bar{\Omega} })} =e^{\Omega} e^{\bar{\Omega} } $ and the gauge transformation
 $ e^{\Omega} e^{\bar{\Omega} } \mapsto e^{-i\bar{\Lambda}} e^{\Omega} e^{\bar{\Omega} } e^{i\Lambda} $ from (\ref{gauge5}) and (\ref{gauge3}) . 
Then, it is reasonable to identify $e^{\Omega}e^{\bar{\Omega}}$ with $e^V$.
Through this identification, we can derive the following correspondences:
\begin{align}
\begin{array}{ccc}
 &\text{covariant approach} & \text{conventional approach}  \\
\text{covariant derivative} & \nabla_A & \nabla_{cA}=e^{-\bar{\Omega}} \nabla_A e^{\bar{\Omega}}  \\
 &e^{-\Omega} D_\alpha e^{\Omega} &  e^{-V}D_\alpha e^V  \\
& e^{\bar{\Omega}} \bar{D}_{\dot{\alpha}} e^{-\bar{\Omega}} &  \bar{D}_{\dot{\alpha}} \\
&  -i \{ \nabla_\alpha , \bar{\nabla}_{\dot{\alpha}} \}    &  -i \{ \nabla_{c\alpha}, \bar{\nabla}_{c\dot{\alpha}} \}  \\
&   & \nabla_{aA}=e^{\Omega} \nabla_A e^{-\Omega} \\
 & &  D_\alpha\\
 && e^V \bar{D}_{\dot{\alpha}} e^{-V} \\
&& -i \{ \nabla_{a\alpha}, \bar{\nabla}_{a\dot{\alpha}} \} \\
\text{field strength} & W_\alpha & W_{0\alpha} = e^{-\bar{\Omega}} W_\alpha e^{\bar{\Omega}} \\
& \bar{W}_{\dot{\alpha}}& \bar{W}_{0\dot{\alpha}}= e^\Omega \bar{W}_{\dot{\alpha}} e^{-\Omega} \\
\text{chiral adjoint matter} & \Phi & \Phi_0 = e^{-\bar{\Omega}} \Phi e^{\bar{\Omega}}   \\
 & (\bar{\nabla}_{\dot{\alpha}} \Phi =0 ) &( \bar{\nabla}_{c\dot{\alpha}} \Phi_0= \bar{D}_{\dot{\alpha}} \Phi_0=0 )\\
\text{anti-chiral adjoint matter} & \bar{\Phi} & \bar{\Phi}_0= e^\Omega \bar{\Phi} e^{-\Omega} 
\end{array}
\label{relation}
\end{align}
Using these correspondences, we can map the action (\ref{action5}) to the action (\ref{action}), and we thus find that these two approaches give the same theory.

Here we also list their gauge transformations:
\begin{align}
\begin{array}{ccc}
 &\text{covariant approach} & \text{conventional approach}  \\
\text{generator} & K & \Lambda, ~\bar{\Lambda} \\
 & (\bar{K} = K) & (\bar{D}_{\dot{\alpha} } \Lambda =0,~D_\alpha \bar{\Lambda}=0) \\
\text{covariant derivative} & \nabla'_A=e^{-iK} \nabla_A e^{iK}
&\nabla'_{cA} = e^{-i \Lambda} \nabla_{cA} e^{i \Lambda} \\
&& \nabla'_{aA} =  e^{-i \bar{\Lambda} } \nabla_{aA}e^{i \bar{ \Lambda}}\\
\text{chiral adjoint matter} & \Phi' =e^{-iK}\Phi e^{iK}
& \Phi'_0 = e^{-i \Lambda} \Phi_0 e^{i \Lambda} \\
\text{anti-chiral adjoint matter}  & \bar{\Phi}' =e^{-iK}\bar{\Phi} e^{iK}
& \bar{\Phi}'_0 = e^{-i \bar{\Lambda} } \bar{\Phi}_0 e^{i \bar{ \Lambda}} \\
\text{gauge multiplet} & e^{\Omega'} = e^{-i \bar{\Lambda} } e^{\Omega} e^{iK}
 & e^{V'} = e^{-i \bar{\Lambda} } e^V e^{i \Lambda} \\
 &  e^{\bar{\Omega}'} =  e^{-iK} e^{\Omega} e^{i \Lambda }&
\end{array}
\label{transformation}
\end{align}

\section{Non-Commutative Superspace and $GL(2|2,R)$ Matrix}
\setcounter{equation}{0}
\label{a3}
In this appendix, we present the explicit forms of the matrices $\hat{\theta}^2$, $\hat{\theta}^\alpha \bar{\hat{\theta}}^{\dot{\alpha}}$, $\dots$, $\hat{\theta}^2 \bar{\hat{\theta}}^2$.
From these expressions, we understand that they satisfy the proper (anti-) commutation relations and that these 15 matrices and $1_4$ are linearly independent.
We also confirm that the supertraces, defined by (\ref{supertrace5}), of these matrices are zero, except that of $\hat{\theta}^2 \bar{\hat{\theta}}^2 $, which is consistent with the Grassmann integral $\int d^2 \theta d^2 \bar{\theta}$.


\begin{align}
\hat\theta^1 = \sqrt\gamma 
\begin{pmatrix}
0&0&0&1\\
0&0&0&0\\
0&-1&0&0\\
0&0&0&0
\end{pmatrix}
 ,~\hat\theta^2 =  \sqrt{\gamma}
\begin{pmatrix}
0&0&1&0\\
0&0&0&0\\
0&0&0&0\\
0&1&0&0
\end{pmatrix}
\nonumber,  \\
\bar{\hat{\theta}}^{\dot 1} = \sqrt{\gamma}
\begin{pmatrix}
0&0&0&0\\
0&0&-1&0\\
0&0&0&0\\
1&0&0&0
\end{pmatrix}
,~\bar{\hat{\theta}}^{\dot 2} = \sqrt{\gamma}
\begin{pmatrix}
0&0&0&0\\
0&0&0&1\\
1&0&0&0\\
0&0&0&0
\end{pmatrix}.
\end{align}


\begin{align}
\hat\theta^1 \hat\theta^2 - \hat\theta^2 \hat\theta^1 = 2\gamma 
\begin{pmatrix}
0&1&0&0\\
0&0&0&0\\
0&0&0&0\\
0&0&0&0
\end{pmatrix}
 ,~\bar{\hat{\theta}}^{\dot 1}\bar{\hat{\theta}}^{\dot 2}-\bar{\hat{\theta}}^{\dot 2}  \bar{\hat{\theta}}^{\dot 1} =  2\gamma
\begin{pmatrix}
0&0&0&0\\
-1&0&0&0\\
0&0&0&0\\
0&0&0&0
\end{pmatrix}
\nonumber,  \\
\hat\theta^1\bar{\hat{\theta}}^{\dot 1} - \bar{\hat{\theta}}^{\dot 1}\hat\theta^1 = \gamma
\begin{pmatrix}
1&0&0&0\\
0&-1&0&0\\
0&0&1&0\\
0&0&0&-1
\end{pmatrix}
,~\hat\theta^1 \bar{\hat{\theta}}^{\dot 2} -\bar{\hat{\theta}}^{\dot 2}\hat\theta^1 = 2\gamma
\begin{pmatrix}
0&0&0&0\\
0&0&0&0\\
0&0&0&-1\\
0&0&0&0
\end{pmatrix},\nonumber \\
\hat\theta^2 \bar{\hat{\theta}}^{\dot 2} - \bar{\hat{\theta}}^{\dot 2}\hat\theta^2 = \gamma
\begin{pmatrix}
1&0&0&0\\
0&-1&0&0\\
0&0&-1&0\\
0&0&0&1
\end{pmatrix}
,~\hat\theta^2 \bar{\hat{\theta}}^{\dot 1} -\bar{\hat{\theta}}^{\dot 1}\hat\theta^2 = 2\gamma
\begin{pmatrix}
0&0&0&0\\
0&0&0&0\\
0&0&0&0\\
0&0&-1&0
\end{pmatrix}.
\end{align}


\begin{align}
\{[\hat\theta^1,\hat\theta^2],\bar{\hat{\theta}}^{\dot 1} \} = 2\gamma^{\frac{3}2} 
\begin{pmatrix}
0&0&-1&0\\
0&0&0&0\\
0&0&0&0\\
0&1&0&0
\end{pmatrix}
 ,~\{[\hat\theta^1,\hat\theta^2],\bar{\hat{\theta}}^{\dot 2} \} = 2\gamma^{\frac{3}2} 
\begin{pmatrix}
0&0&0&1\\
0&0&0&0\\
0&1&0&0\\
0&0&0&0
\end{pmatrix}
\nonumber,  \\
\{[\bar{\hat{\theta}}^{\dot 1}, \bar{\hat{\theta}}^{\dot 2} ],\hat\theta^1 \} =  2\gamma^{\frac{3}2}
\begin{pmatrix}
0&0&0&0\\
0&0&0&-1\\
1&0&0&0\\
0&0&0&0
\end{pmatrix}
,~\{[\bar{\hat{\theta}}^{\dot 1}, \bar{\hat{\theta}}^{\dot 2} ],\hat\theta^2 \} =  2\gamma^{\frac{3}2}
\begin{pmatrix}
0&0&0&0\\
0&0&-1&0\\
0&0&0&0\\
-1&0&0&0
\end{pmatrix}.
\end{align}

\begin{align}
[ \hat\theta^1 ,  \{ \hat\theta^2 , [\bar{\hat{\theta}}^{\dot 1}, \bar{\hat{\theta}}^{\dot 2} ] \} ] =  2\gamma^2
\begin{pmatrix}
-1&0&0&0\\
0&-1&0&0\\
0&0&1&0\\
0&0&0&1
\end{pmatrix}.
\end{align}

\end{document}